\documentclass[final,12pt,3p]{elsarticle}

\usepackage{amssymb}

\usepackage{booktabs}
\usepackage{longtable}
\usepackage{lscape}
\usepackage{lineno,hyperref}
\modulolinenumbers[5]
\usepackage{amssymb}
\usepackage{amsmath,eurosym}
\usepackage{gensymb}
\usepackage{calrsfs}
\usepackage{bbding} 
\usepackage{rotating}
\usepackage{tabulary}   
\usepackage{tabularx}
\usepackage{adjustbox}
\usepackage{graphicx}
\usepackage{multirow}
\usepackage{float}
\usepackage{multicol}
\usepackage{nomencl}
\usepackage[table]{xcolor}

\biboptions{comma,square,sort&compress}

\usepackage{caption}
\usepackage{subcaption}

\usepackage{todonotes}
\newcommand{\tinytodo}[2][]{\todo[caption={#2}, size=\small, #1]{\renewcommand{\baselinestretch}{0.5}\selectfont#2\par}}
\newcolumntype{M}[1]{>{\centering\arraybackslash}m{#1}}
\newcommand\hlr[1]{%
  \bgroup
  \hskip0pt\color{red!80!black}%
  #1%
  \egroup}
 \renewcommand{\tinytodo}[2][]{}
 \renewcommand\hlr[1]{#1}

\newcolumntype{M}[1]{>{\centering\arraybackslash}m{#1}}
\newcommand\hlrrevnew[1]{%
  \bgroup
  \hskip0pt\color{red!80!black}%
  #1%
  \egroup}
 
\usepackage{marginnote}

\journal{Renewable and Sustainable Energy Reviews}

\begin{document}

\begin{frontmatter}


\title{Data-driven Predictive Control for Unlocking Building Energy Flexibility: A Review}

\author[label1,label2]{Anjukan Kathirgamanathan\corref{cor1}\fnref{label3}}
\address[label1]{School of Mechanical and Materials Engineering, University College Dublin}
\address[label2]{UCD Energy Institute, O'Brien Centre for Science, University College Dublin\fnref{label4}}

\cortext[cor1]{Corresponding author}

\ead{anjukan.kathirgamanathan@ucdconnect.ie}

\author[label1,label2]{Mattia De Rosa}

\author[label2,label5]{Eleni Mangina}
\address[label5]{School of Computer Science, University College Dublin}

\author[label1,label2]{Donal P. Finn}


\begin{abstract}
Managing supply and demand in the electricity grid is becoming more challenging due to the increasing penetration of variable renewable energy sources. As significant end-use consumers, and through better grid integration, buildings are expected to play an expanding role in the future smart grid. Predictive control allows buildings to better harness available energy flexibility from the building passive thermal mass. However, due to the  heterogeneous nature of the building stock, developing computationally tractable control-oriented models, which adequately represent the complex and nonlinear thermal-dynamics of individual buildings, is proving to be a major hurdle. Data-driven predictive control, coupled with the ``Internet of Things", holds the promise for a scalable and transferrable approach, with data-driven models replacing traditional physics-based models. This review examines recent work utilising data-driven predictive control for demand side management application with a special focus on the nexus of model development and control integration, which to date, previous reviews have not addressed. Further topics examined include the practical requirements for harnessing passive thermal mass and the issue of feature selection. Current research gaps are outlined and future research pathways are suggested to identify the most promising data-driven predictive control techniques for grid integration of buildings. 
\newline
\newline
\textit{Highlights:}
\begin{itemize}
    \item A review of 115 data-driven predictive control studies in building energy management.
    \item Growing trend of research on data-driven predictive control in recent years.
    \item High level of simplification in modelling seen, mostly applied to single buildings.
    \item Gaps identified were feature selection, benchmarking, data quality and scalability.
\end{itemize}

\end{abstract}


\end{frontmatter}


\section{Introduction}
\label{sec1}
\subsection{The Importance of Building Energy Flexibility in the Smart Grid}
\label{sec1.1}
Renewable energy sources such as wind and solar are intrinsically variable by nature and this creates a stability issue for the grid with the fluctuating supply needing to be balanced with demand \cite{Lund2015}. In their review of the future of the low-carbon electricity grid, \citet{Greenblatt2017} review such conventional renewable energy technologies. They suggest that these technologies alone do not represent an ideal solution and that grid integration is required in conjunction to deliver a reliable and robust electricity system. \citet{Holttinen2009} focus on the impacts of large amounts of wind power on the design and operation of power systems. Similarly, \citet{Haegel2017} address the same question and barriers with large penetrations of solar photovoltaics. \citet{Villar2018} summarise some of the challenges faced by this new power system and the need for new flexibility products and markets. The flexibility to manage any mismatch can come from either the supply side (through the use of dedicated conventional power plants or storage) or from the demand side \cite{Lund2015, Cochran2014}. Demand Side Management (DSM) is one such grid integration strategy and can be broadly categorised as actions that influence the quantity, patterns of use or the primary source of energy consumed by end users \cite{Hull2012}. Demand Response (DR) is one promising facet of DSM where consumers curtail or shift their electricity usage in response to financial or other incentives. Within DR, there are different strategies based on the response times, services offered and business models \cite{Klein2017}.

With buildings representing about 40\% of the total primary energy consumption in Europe \cite{Economidou2011}, they are very relevant to participation in DR and the provision of energy flexibility. Further, the thermal mass of buildings allows them to be used as a thermal energy storage making them potentially very useful in DSM \cite{Reynders2013}. Commercial buildings are of particular interest given their greater thermal mass and common usage of space conditioning. This is often through the use of heating, ventilation and air conditioning (HVAC) systems and this HVAC load is one such load that can be shifted using the thermal mass of the building. These HVAC systems are often integrated with Building Automation Systems (BAS) or Building Energy Management Systems (BEMS) which can be used to automate DR measures. These systems are also capable of receiving signals directly from the electricity grid \cite{Hao2012}. Buildings may often also possess active thermal storage, active electric storage (batteries), indirect electric storage (Electric Vehicles (EVs)) and on-site generation as further sources of energy flexibility.

DR programs can be categorised as being either price or incentive based \cite{Vardakas2015}. Incentive-based programmes pay customers (end-users) to shift their electricity consumption at times requested by grid operators. Grid operators can be either Distribution System Operators (DSO), who are generally responsible for the operation of the low-voltage distribution system and delivery of power to the end consumers; or Transmission System Operators (TSO), who are responsible for the operation of the high-voltage transmission system and ensuring its stability. Generally, a building is required to be capable of meeting a minimum required reduction in power consumption. While individual buildings may not be capable of meeting this reduction, aggregators are a market actor who contract with these buildings and combine the available power reduction and offer this to the grid operator, receiving a percentage of the value to the grid operator created by applying the DSM measure \cite{Jensen2017}. Given that energy flexibility is a resource that is aggregated from many buildings, which are all unique in the way that they are constructed, designed and operated, a scalable and transferrable method of assessing and harnessing this energy flexibility is required.

\subsection{Approaches to Assess and Harness Building Energy Flexibility}
\label{sec1.2}

Approaches to assess and harness energy flexibility generally require as input; data, a model and a control framework. For example, a model capturing the thermal dynamics of the building and its heating or cooling system is often required to ensure that the thermal comfort of the occupants is not compromised due to shifting the HVAC load for DSM purposes \cite{Biyik2019}. A control framework is required to actuate the building energy systems appropriately to ensure that the performance objectives of the energy systems are met \cite{Ascione2016}. 

Although the optimal control of a HVAC system is a complex multi-variable problem, there has been increased focus in literature recently for control strategies such as Model Predictive Control (MPC) in buildings. However, the standard control seen in most buildings today is simple Rule-Based Control (RBC) \cite{MaasoumyMehdi;Sangiovanni-Vincentelli2015}. RBC is often outperformed by MPC which is predictive in nature and hence able to take future disturbances (e.g., grid or weather signal) into account. Moreover, MPC can harness the building thermal storage characteristics by predicting the building thermal behaviour evolution \cite{DeConinck2015, Gwerder2013}. MPC control requires a linear model (convex optimisation problem) if an unique solution is to be guaranteed - although some studies have considered highly nonlinear building models for MPC as well \cite{Nghiem2017, Afram2017}. A fully linear model is often not feasible with buildings, especially if they include ventilation models where bi-linearities are introduced due to modelling the product of temperature and mass flow rate \cite{Sturzenegger2014}. The challenges of generating a suitable and accurate model are one of the most significant limitations of MPC and have led to MPC largely being constrained to the research field to date \cite{Sturzenegger2016, Henze2013}. In fact, in their study of implementing MPC in a typical Swiss office building, \citet{Sturzenegger2016} concluded that deploying MPC had too high a cost relative to the operating cost savings given the energy prices at that time and tools available for model development.

The challenges around the implementation of MPC have led researchers to investigate data-driven approaches where the predictive capability of MPC is retained without the expense of first principle based model generation. One of the challenges with black-box models is that they are often unsuitable for integration with control as the models are generally highly non-linear which adds additional computational complexity to the MPC problem \cite{Serale2018}. Several techniques have been used to mitigate these issues, e.g., using separation of manipulated control and non-manipulated disturbance variables \cite{Behl2016} and using branch and bound techniques on a discrete search space \cite{Ferreira2012}. A few terms have appeared in the literature for the model and control method, where the controller is designed by directly using online or offline input/output data from the controlled system rather than starting from a first principles based model. \citet{Jain2017} coined the term `DPC' (Data Predictive Control) in their work whereas other literature uses the term `DDC' (Data-Driven Control) \cite{Hou2016}. As the field of treating buildings as Cyber-Physical Systems (CPS) is a very new area of research, there has been little research in terms of comparing these different techniques (apart from \cite{Schmidt2018} in terms of energy efficiency).

\subsection{Previous Reviews}
\label{sec1.3}

There have been several review papers that have looked at the various data-driven approaches employed in building energy management. \citet{Amasyali2018} and \citet{Wei2018} provide review papers that look solely at the problem of predicting electricity consumption. \citet{Deb2017} provides a similar review with the problem treated as a time series forecasting one. A state of the art review on data-driven modeling of building thermal dynamics is provided by \citet{Wang2019}. Whilst these studies did not consider the control problem, \citet{Hou2016} provide a survey and classification of `DDC', although the focus was not on the building domain. The review of \citet{Maddalena2020} investigates the HVAC control problem with a focus on data-driven models. They define the main aspects of the building climate problem, summarised succinctly as; large sampling periods, low-quality measurements, operations-critical (operating envelopes constrained by building occupancy and thermal comfort requirements) and difficult modelling. There have been several review papers focusing exclusively on MPC for building energy management applications; studies looking at the theory and applications of MPC in building HVAC systems \cite{Afram2014}, a review with regards to presenting a unified framework \cite{Serale2018} and several works reviewing building control systems and modelling techniques from a broader perspective \cite{Afroz2017, Hameed2014, Li2014}. \citet{Hameed2014} provide a review of optimised control systems for energy management in buildings. However, these reviews do not consider the applicability of such control systems for meeting the energy flexibility needs of the smart grid. Those that have are not focused on data-driven approaches, e.g., the review of assessment and control of DR programmes in a residential context \cite{Pallonetto2020} or are focused on a particular technique only, such as the work of \citet{Vazquez-Canteli2019} on Reinforcement Learning. 

\subsection{Motivation}
\label{sec1.4}
Given that the generation and identification of a suitable control-oriented model of the thermal dynamics of a building is considered to be one of the most significant hurdles in the implementation of predictive control in building energy systems, this is also one of the biggest limitations in unlocking energy flexibility potential in building stock. This justifies a literature review of existing data-driven approaches, with this paper aiming to address the following research gaps:
\begin{enumerate}
    \item Although different data-driven predictive control techniques have emerged in recent years, no systematic evaluation of the applicability or the suitability of any of these techniques to address the DSM problem in buildings has been carried out to date.
    \item The practical requirements of the building passive thermal mass control-oriented model for harnessing the inherent energy flexibility continues to be not well understood.
    \item The type of input features used for training data-driven modelling and control frameworks for DSM in buildings has received little attention in the literature.
\end{enumerate}
With these research gaps in mind, this review provides the following contributions and novelty by:
\begin{enumerate}
    \item Giving a holistic overview and categorisation of the variety of data-driven predictive approaches used in building energy management and DSM, with a focus on the building passive thermal mass and the modelling of the associated dynamics, which from a subject perspective is novel in a survey.
    \item Providing insights on the coupling of system models and control integration by means of a qualitative analysis of the approaches used and their suitability in the context of a building energy flexibility problem, which to date has not been addressed in literature.
    \item Assessing emerging techniques to address the challenge of feature selection in the context of the availability of ubiquitous data from building energy management systems, where a paucity of reviews exist to date.
\end{enumerate}
The broader application of this study is understanding the data-driven techniques most suitable for predictive control in building energy systems, whilst being scalable and transferable between buildings. To give a few examples of potential applications, for the end-user (i.e., building energy manager and/or operator), such applications may facilitate participation in DR programmes, supporting energy efficiency efforts and reducing the carbon footprint of the building. For aggregators, possible applications include the facilitation of direct control of their building portfolio for DSM and assessing the flexibility potential of their portfolio.

The paper is organised as follows: Section \ref{sec2} provides essential background to the concept of building energy flexibility, and identifies the models and control techniques used for building energy management. Section \ref{sec3} presents the methodology used in the current review to compile the studies that are investigated and to categorise the studies. Section \ref{sec4} presents a classification table of the reviewed studies summarising key information such as the type of control, type of model, application of the study with a particular focus on energy flexibility applications. Visualisations are also provided illustrating trends evident from the review and further analysis is provided where relevant. Section \ref{sec5} provides a discussion of the key findings, Section \ref{sec6} gives the conclusions and Section \ref{sec7} summarises potential future research directions to enable and enhance the further use of data-driven predictive control for building energy flexibility applications.

\section{Background}
\label{sec2}

\subsection{Demand Side Management and Building Energy Flexibility}
\label{sec2.1}

In the scope of this review, it is first useful to provide a formal definition of building energy flexibility. The International Energy Agency (IEA) Energy in Buildings and Communities Program (EBC) Annex 67 is dedicated to the issue of building energy flexibility \cite{Jensen2017}. The Annex gives the definition as \textit{``The Energy Flexibility of a building is the ability to manage its demand and generation according to local climate conditions, user needs, and energy network requirements. Energy Flexibility of buildings will thus allow for DSM/load control and thereby DR based on the requirements of the surrounding energy networks"} \cite{Jensen2017}. From this definition, it is clear that any approach to assess and harness building energy flexibility needs to be able to capture both the energy demand and generation potential (if it exists) of a building, as well as some form of communication and/or coordination with the aggregator or electricity grid operator to ascertain the needs of the network. Note that per the three distinct flexibility products that \citet{Villar2018} describe in their review, our definition of building energy flexibility is taken to cover both ramping capacity (power) and energy products.  

The energy flexibility from buildings may ultimately be used for different purposes, including: power balance for frequency control by the TSO, congestion management by either the TSO or DSO or portfolio energy balancing by a Balancing Responsible Party (BRP). For further details on these relevant stakeholders, the reader is referred to the review of \citet{Villar2018}. This demand side flexibility can be encouraged through various mechanisms such as market prices, distribution tariffs, reliability signals and incentive payments from aggregators who sell on the aggregated flexibility on the markets. For example, in the day-ahead market, electricity suppliers are required to nominate their electricity bids to meet forecasted supply and demand, and on this basis, a day-ahead price schedule is set. Real-Time Prices (RTP) and spot prices are other examples of market pricing schemes and are dynamic and better reflect the more granular variations (both temporal and locational) in the balancing requirements of the grid at the intra-day level. Whilst these pricing signals may come from the wholesale market, suppliers or aggregators often pass on such signals to end-use stakeholders (customers), such as buildings, at the retail level. Regulation signals require a building to track a certain power consumption profile and this is a type of signal commonly used for ramping capacity (power). Such ancillary services are often found at the wholesale market level. For a more comprehensive summary of flexibility markets and products, the reader is referred to the review by \citet{Villar2018}. Furthermore, the reader is referred to \citet{Pallonetto2020} for a summary and classification of DR programmes, in the context of residential sector applications.

\subsection{Building Modelling Frameworks}
\label{sec2.4}

Models of building components, systems and sub-systems are required to predict the whole building and sub-system behaviour, such as their energy consumption and evolution of internal temperature, humidity and air quality \cite{Li2014}. The work of Maasoumy provides a review of many of the different building simulation and modelling approaches employed in smart building energy management \cite{MaasoumyMehdi;Sangiovanni-Vincentelli2015}. This review was focused on white-box and grey-box approaches while black-box approaches were not discussed. The summary below provides a brief overview of white-box, grey-box and black-box (with the latter two being data-driven per the definition of this review) approaches.

\begin{itemize}
\item \textbf{White-box models}: These are based on physical and first-principle based modelling. Models can be high-fidelity, produced from simulation software such as EnergyPlus, IESVE and TRNSYS to name a few commercially available products, to reduced order models \cite{Li2014}. Models can be based on static equations, or more commonly, dynamic equations which represent the heat balance time evolution \cite{Amara2015}. Generally, and for detailed white-box simulation models in particular, complete physical knowledge (including geometry, material properties, construction, etc.) of the building is required to build such models, and this is often one of the biggest drawbacks of such approaches \cite{Amara2015}. Detailed white-box simulation models, although often able to model the dynamics in a comprehensive manner, are also generally considered not to be suitable for on-line control of buildings due to their large computational overhead \cite{Li2014, Lehmann2013, Serale2018}. Nevertheless, there have been a few successful implementations integrating such models with control \cite{Aftab2017, Ascione2016, Gomez-Romero2019, Kontes2018}, although these are considered to be out of scope for this review. Models of systems such as thermal energy storage and electric batteries are often simpler and well established and hence physics-based models of these systems are commonly employed \cite{Mirakhorli2018, Vrettos2013}. These models are not considered within the scope of this review.

\item \textbf{Black-box models}: Black-box approaches completely ignore the physics and are fully empirical models \cite{Amara2015}. These approaches require extensive datasets including HVAC power consumption, indoor and outdoor temperatures and setpoints to train the model. The training datasets need to be large (covering all seasons) and rich (covering all possible operational envelopes) \cite{Li2014, Li2015}. This is often the most challenging aspect of developing data-driven models for building energy management. Detailed simulation models, if they exist for the building, can provide this training data but these often fail to capture the stochastic nature of real data. Given large datasets with many features (variables or sensors in this case), feature assessment and selection becomes a critical process to develop an accurate yet efficient model. \citet{Kathirgamanathan2019} elaborates on the importance of feature selection in data-driven models for harnessing building energy flexibility. Models used to estimate DR potential range from data-driven regression based models \cite{Yin2016, Kara2014} and decision trees \cite{Behl2016} to completely black-box models such as Neural Networks and Support Vector Machines \cite{Kapetanakis2017}.

\item \textbf{Grey-box models}: These are a mix of physics-based and empirical based models. Resistance-Capacitance (RC) (or Lumped-Capacitance) models are one example of such models where the parameters are identified using experimental data. Here a set of continuous time (stochastic) differential equations are derived and system identification techniques are used for parameter estimation. \citet{Privara2012} presents an approach to develop and select a model suitable for predictive control focusing on grey-box models only. \citet{Privara2013} further concluded that grey-box models are ideal for buildings that are not overly complex and recommended subspace methods. Commonly used, a state-space (SS) model is a set of input, output and state variables represented by first order differential equations. Literature often provides the benefits of grey-box models as allowing physical interpretation of the parameters of the model and an understanding of some of the underlying physical phenomena behind the building behaviour \cite{Wang2019, Bacher2011}.
\end{itemize}

\subsection{Building Control Frameworks}
\label{sec2.5}
The control of the energy systems in a building is a unique and complex problem given nonlinear dynamics of the building, time-varying disturbances and constraints, interacting (and possibly conflicting) control loops within building systems and traditionally poor availability of data from these systems \cite{Afram2014}. For this reason, it is prudent to review the control techniques used in the building energy management domain with rigour and consideration to the above mentioned challenges. Given a desire for buildings to be ``grid-interactive", the controller is further required to be able to respond to external signals representing the needs of the grid. In this vein, treating buildings as CPS has received increased attention in recent years \cite{Schmidt2018, Behl2016a}. CPS are ``integrations" of computation and physical processes. Embedded computers and networks monitor and control the physical processes, usually with feed-back loops where physical processes affect computations and vice-versa \cite{Lee2008}. The review by \citet{Schmidt2018} of buildings as CPS focused on the application of building energy efficiency and reducing operational energy consumption. \citet{MaasoumyMehdi;Sangiovanni-Vincentelli2015} provides a comprehensive introduction to building control design and the reader is pointed to this reference for further details.

Note that the lines demarcating control techniques and models are becoming blurred with techniques such as neural network (NN) control where the black box model is trained on controller input-output attempting to replace the controller. Control can further be classified as model-based (either physics based or data-driven) or model-free (e.g., Reinforcement Learning). The main broad classes of control used to categorise the studies reviewed are detailed further below.

\subsubsection{Rule-Based Control (RBC)}
\label{sec2.5.1}
Due to its simplicity, the standard practice in the building automation industry is RBC at the supervisory level, essentially a set of if-then-else rules used to determine the best operating points for the system \cite{Maddalena2020}. These commonly include on/off controllers and PID controllers (classical controllers) at the device level. Generally, the efficacy of these rules depends on the expertise and knowledge of the building operator \cite{Rockett2017}. RBC is able to exploit energy flexibility if the appropriate decision criteria is used in the rules. There are a few such examples of RBC in the literature incorporating predictive models \cite{Gwerder2013, Klein2017, Macarulla2017, Manandhar2015}. However, these approaches are increasingly being found to be inadequate with meeting complex objectives, especially as the system complexity increases and they require significant human oversight \cite{Rockett2017}.

\subsubsection{Model Predictive Control (MPC) \& Variants}
\label{sec2.5.2}

MPC is based on the solution of an optimal control problem on an iterative basis for a finite horizon (therefore also called receding horizon control). For every time step, the optimal control sequence is found for a finite time horizon whilst meeting constraints. The method is robust against model errors and the effect of future exogenous disturbances. This is because only the first step of the optimal control strategy is implemented and at the next time step a new plant measurement is taken and the process is repeated for the new horizon \cite{MaasoumyMehdi;Sangiovanni-Vincentelli2015}. MPC is theoretically very suitable for optimal control of building. MPC provides a framework for the optimisation of a cost function (e.g., cost, energy consumption or CO$_2$ emissions) subject to constraints over the prediction horizon (e.g., thermal comfort constraints, HVAC system limits) \cite{Rockett2017, Serale2018}. Furthermore, MPC is very suitable for utilising passive or active thermal or electrical storage for shifting energy consumption as a form of DSM. This is because of the ability of MPC to utilise a model to predict the evolution of the state variable (e.g., zone temperature) under the influence of the controlled inputs coupled with predictions of future disturbances. These disturbances may be a signal representing the flexibility required from the grid, or exogenous disturbances, such as ambient temperature or other external weather conditions which influence the energy demands of a building. 
\par
Research into MPC applications in buildings in the academic community has surged in the last decade as Figure \ref{fig:2}(b) in this article and the review of \citet{Serale2018} shows. A concise summary is presented below of the main tenets of the MPC framework. The reader is referred to \cite{Serale2018} for a detailed description of terminology and taxonomy with respect to MPC and an attempt at describing the MPC formulation framework with respect to energy efficiency control of buildings and HVAC systems.

A control-oriented dynamical model of the system is required to capture the behaviour of the system under controlled and uncontrolled inputs. These models can either be physics-based (classical MPC) or data-driven (sometimes called DPC or DDPC in the literature) as outlined in Section \ref{sec2.4}. As per Section \ref{sec1.2}, the greatest challenge for classical MPC approaches is deriving a suitable model of the building thermal dynamics that is compatible with the optimal control problem and yet captures the relevant dynamics of the building thermal behaviour \cite{Sturzenegger2016, Henze2013}. The need for a compatible model is also faced by data-driven MPC approaches, which exhibit a particular challenge to ensure the convexity of the optimal control problem. A further challenge is ensuring that the data-driven model is capable of capturing the dynamics and operating ranges of the building not seen in the training data. Comparisons between the two approaches are rare for large complex buildings, but notable studies include the work of \cite{Jain2017} and \cite{Smarra2018}. In both studies, the authors compared DPC with classical MPC for a bilinear building model, where they found a comparable performance (cost minimisation) of DPC to MPC, whilst bypassing the need for physical modelling of the building. Where the system controlled is complex, distributed MPC problems can be employed \cite{Mirakhorli2018}. 

Further models are required to predict the evolution of uncontrolled disturbance variables such as external weather conditions and energy prices (hereby referred to as disturbance models). A further model is necessary to test and validate an MPC framework at the design stage \cite{Serale2018} (hereby referred to as the surrogate simulation model). Often, this is a high-fidelity simulation building model which closes the control loop in simulation and provides the feedback to the controller.

In terms of the optimal control problem itself, the objective function is a crucial element and dictates the overall global objectives of the MPC framework. Generally, this will be dictated by the relevant stakeholder preferences. An economic controller, unlike a controller purely minimising energy consumption, aims to reduce the cost of energy consumption and given time-varying prices can exploit the energy flexibility of a building. Comfort related objective functions are common, minimising discomfort to occupants and are often measured as deviations from a target temperature setpoint. Multi-objective functions can take multiple objectives into account with the respective weighting towards the individual aspects dictated by the stakeholders or building managers. As explained in \citet{Serale2018}, objective functions are generally either quadratic (common in tracking problems), linear or min/max (e.g., reducing peak power).

There are variations to the classical MPC formulation. In most practical examples, the controlled system may be influenced by uncertain disturbances, leading to deviations of the state variable (e.g., building indoor zonal temperature) from the expected trajectory. Whilst deterministic MPC algorithms are unable to take this into account, and hence vulnerable to degradations of control performance or violations of constraints, Stochastic Model Predictive Control (SMPC) introduces chance constraints \cite{Shang2019}. The introduced probabilistic distributions of disturbances are often not exactly known, making the optimisation problem computationally difficult to solve. Robust Model Predictive Control (RMPC) utilises a bounded set to describe the uncertainty with the worst case scenario considered. This ultimately leads to over-conservatism and is a weakness of RMPC \cite{Shang2019}.

\subsubsection{Reinforcement Learning}
\label{sec2.5.3}

Reinforcement learning (RL) is an area of machine learning, where an agent learns to take the optimal set of actions through interaction in a dynamic environment (such as a building subject to changing weather conditions, varying grid requirements and occupants with thermal comfort requirements), with the goal of maximising a certain reward quantity \cite{Mason2019}. For a comprehensive introduction to this field, the reader is referred to standard textbooks \cite{Sutton2014}. The major advantage of this method is that it is potentially a model-free control approach avoiding the need to create non-generalisable models of the building. Q-learning, where the state-action value function is learned, is an established RL method and commonly found in literature \cite{Costanzo2016}. In standard RL, the policy (based upon which the agent takes action) is updated online at every time step, however, Batch Reinforcement Learning (BLR) is a variation where the policy is calculated offline using a batch of historical data \cite{Costanzo2016}. Another relatively recent variation is the use of deep neural networks in conjunction with RL with the neural network relating the value estimates and state-action pairs \cite{Wei2017}. The reader is referred to the review conducted by \citet{Vazquez-Canteli2019} for a comprehensive review of the use of RL in DR applications.

\section{Methods}
\label{sec3}

This literature review summarises studies (with publication dates from 2010 to a cut-off date of 01/11/19) utilising data-driven predictive control (i.e., capable to react based on forecasted variables such as environmental, grid-signals or occupancy) for building energy management and DSM. Only articles that consider the modelling of the building thermal dynamics (but including model-free studies such as RL harnessing building passive thermal mass) are included. Hence, studies that consider energy flexibility arising purely from active electric storage, active thermal storage or DERs such as solar PV, are disregarded. For these systems, often simple and well understood physics-based equations are suitable as control-oriented models. Accordingly, they do not necessarily present the same challenges with model identification that building thermal dynamics presents. All models and control techniques that use any form of sensor data from the building in question (either real and measured or synthetic and simulated) are considered data-driven approaches. The articles were screened, such that only those related to the following applications were selected:
\begin{itemize}
\item DSM (incl. harnessing energy flexibility, grid coordination, peak power reduction, increasing self-consumption);
\item Energy efficiency;
\item Cost savings;
\item Maintaining or improving thermal comfort;
\item MPC development (incl. MPC alternatives);
\item Co-simulation development.
\end{itemize}

In order not to limit the number of studies considered, the scope of the review was extended to those that utilised predictive control for applications such as improving thermal comfort/energy efficiency and economic management, rather than limiting it to those for energy flexibility applications only. The rationale is that such studies can also be highly relevant for energy flexibility applications, as often a change in objective function in these cases is sufficient for the approach to be used to shift energy consumption with regards to the needs of the grid. Hence, the insights from these studies regarding model and controller development for slightly different applications are highly relevant.

Applying the above selection criteria, 115 studies were selected and these were categorised based on the points of consideration summarised in Table 1. This table outlines the scope of the review and also provides the relevant section for the analysis. Where appropriate, the reason why a topic is considered to be beyond the scope of this review is given.

\begin{table}[h!]
\centering
\caption{Import Aspects Considered in the Review and Outline of Scope.}
\scriptsize
{\setlength\tabcolsep{3.5pt} 
\label{table:1}

\begin{tabular}{p{3.5cm}p{2.7cm}p{9.3cm}}
\toprule
    Important Aspects &              Included in Review? &  Details and Justification where not in Scope\\
\midrule
    Application Domain &            Section \ref{sec4.2} &  Categorisation into: DSM, energy efficiency, cost savings, thermal comfort, MPC and co-simulation. \\
    Building &            Section \ref{sec4.3} &  Categorisation into: commercial, commercial group, institutional, multi zone, residential, residential group, single-zone. \\
    Model &            Section \ref{sec4.4} &  Categorisation of the type of model used for the building passive thermal mass: black box, model-free, reduced order, regression based. \\
    Control Methodology &            Section \ref{sec4.5} &  Categorisation of the type of control used for harnessing building energy flexibility: MPC, MPC approx, hybrid MPC, RBC, RL. \\
    Objective Function &            Section \ref{sec4.6} &  Categorisation of the types of objective functions used in optimal control problems: cost, energy, comfort, other. \\
    Flexibility Resources &            Section \ref{sec4.7} &  Categorisation of the different types of flexibility resources considered together with building passive thermal mass: active thermal, active electric, EV, on-site generation. \\
    Feature Selection &            Section \ref{sec4.8} &  Whether feature selection was applied in the studies and what methods were used. \\
    Grid Signals &            Section \ref{sec4.9} &  Categorisation of what type of grid signals are used by the predictive control frameworks: DR, day-ahead, dynamic, other, regulation, TOU. \\
    Occupancy &            Section \ref{sec4.9} &  Consideration whether occupancy was used as an input for the predictive control. How occupancy is forecast is beyond the scope of this review. See \citet{LiZ2018} and \citet{Kleiminger2014} for further details on this. \\
    Weather &            Section \ref{sec4.9} &  Consideration whether weather forecasts were used as input for the predictive control. The climate type itself was not recorded as this is not relevant to the modelling and control focus of this review. \\
    Control Test-bed &            Section \ref{sec4.10} &  Categorisation of the types of test-bed used for evaluating the controller: real and/or simulation. \\
    Controller Performance &            Section \ref{sec4.11} &  Investigation of the controller performance on a sample of the literature surveyed. This is difficult to compare in a meaningful manner given different dynamic boundary conditions, buildings and market conditions. \\
    Predictive Accuracy &            No &  This is difficult to compare in a meaningful manner given different dynamic boundary conditions, training data and buildings used in the literature. Therefore, this was considered to be beyond the scope of this review. See \cite{Amasyali2018, Wei2018, Privara2013} for further details on this. \\
    Model Uncertainty &            No &  This is difficult to compare in a meaningful manner given different dynamic boundary conditions, training data and buildings used in the literature. Model uncertainty is considered to be beyond the scope for this review. The reader is referred to the reviews of \citet{Tian2018} and \citet{Coakley2014} for discussion on this topic. \\
\bottomrule

\end{tabular}

}
\end{table}
\normalsize

The subject of model prediction accuracy is one such topic excluded from this review. Other review papers have explicitly addressed this issue \cite{Amasyali2018, Wang2019, Privara2013} and these reviews have found that a variety of different metrics have been used to report model accuracy. Further, given different test periods, climates and buildings, a direct comparison of these quantitative values is generally not necessarily meaningful. Similarly, the topic of model uncertainty is also considered to be beyond the scope for this review. The reader is referred to the reviews of \citet{Tian2018} and \citet{Coakley2014} for discussion on this topic. A full list of the reviewed papers and categorisation on the above points is presented in Table 2. Section \ref{sec4} is organised with reference to Table 2, where each section subheading is taken from Table 2. From these results, a discussion is presented and research gaps and potential future research directions were identified.

\section{Data-driven Predictive Control Approaches}
\label{sec4}

\subsection{Review Summary}
\label{sec4.1}

Table 2 provides a summary of the reviewed articles meeting the scope described above and categorises the works in terms of the key factors outlined in Section \ref{sec3} (Methods). The remaining part of this section critically reviews and comments about the individual studies themselves as well as broader trends discovered from this survey.

\begin{landscape}
\scriptsize
{\setlength\tabcolsep{3.5pt} 
\label{table:2}
\begin{longtable}{p{1cm}p{4cm}p{2cm}p{2cm}p{2cm}p{1cm}p{2cm}p{2cm}p{1.2cm}p{1.2cm}p{2cm}p{1.2cm}}
\caption{Application, Building Model, Control Technique and Energy Flexibility Scope of Reviewed Data-Driven Predictive Control Studies,}\\
\toprule
                        Ref & Application & Building Type & Testbed & Control Type & Comfort Metric & Building Model & Building Model Details & Feature Selection & Weather Forecast & Grid Signal & Occupancy Forecast \\
\midrule
\endhead
\midrule
\multicolumn{12}{r}{{Continued on next page}} \\
\midrule
\endfoot

\bottomrule
\endlastfoot
           \cite{Afram2017} &       Energy Efficiency &        Residential &         Simulation &                     MPC &                   &         Black Box &             NN &                No &        Yes &                        TOU &           No \\
           \cite{Allen2018} &  DSM &         Commercial &         Simulation &                     MPC &                   &     Reduced Order &                         SS &                No &        Yes &                       None &          Yes \\
            \cite{Aoun2019} &  DSM &        Residential &         Simulation &                     MPC &                 Temp &     Reduced Order &                    SS (RC) &                No &        Yes &                        TOU &           No \\
            \cite{Behl2016a} &  DSM &         Commercial &         Simulation &                     MPC &                 Temp &         Black Box &                 Tree based &                No &        Yes &                         DR &          Yes \\
            \cite{Behl2016} &       Energy Efficiency &         Commercial &         Simulation &                     MPC &                 Temp &         Black Box &                 Tree based &                No &        Yes &                         DR &          Yes \\
     \cite{Bianchini2019} &            Cost Savings &         Commercial &         Simulation &                     MPC &                   &  Regression Based &          Linear Regression &                No &        Yes &                         DR &          Yes \\
           \cite{Biyik2019} &  DSM &         Commercial &         Simulation &                     MPC &                   &     Reduced Order &                    SS (RC) &                No &        Yes &                        TOU &           No \\
         \cite{Bunning2019} &       Energy Efficiency &        Residential &               Real &                     MPC &                 Temp &         Black Box &                 Tree based &                No &        Yes &                       None &          Yes \\
          \cite{Adrian2019} &       Energy Efficiency &         Commercial &         Simulation &                     MPC &                   &     Reduced Order &                         SS &                No &        Yes &                       None &           No \\
          \cite{Cigler2012} &       Energy Efficiency &         Commercial &         Simulation &                     MPC &                PMV$^2$ &     Reduced Order &                         SS &                No &        Yes &                       None &          Yes \\
          \cite{Cigler2013} &       Energy Efficiency &         Commercial &         Simulation &                     MPC &               Temp$^2$ &     Reduced Order &                         SS &                No &        Yes &                       None &          Yes \\
         \cite{Coetzee2019} &  DSM &        Residential &         Simulation &                     MPC &                   &     Reduced Order &                         SS &                No &        Yes &                       None &          Yes \\
        \cite{Costanzo2016} &            Cost Savings &        Single Zone &  Real + Simulation &  RL &                   &         Black Box &             NN &                No &        Yes &                    Dynamic &           No \\
         \cite{Cotrufo2020} &       Energy Efficiency &         Commercial &               Real &                     MPC &                   &         Black Box &           Gaussian Process &                No &        Yes &                       None &           No \\
             \cite{CuiBorui;FanCheng;Xiao2016} &  DSM &        Residential &         Simulation &                     MPC &                   &     Reduced Order &                    SS (RC) &                No &        Yes &                       None &           No \\
      \cite{DeConinck2013} &  DSM &         Commercial &         Simulation &                     MPC &                   &     Reduced Order &                    SS (RC) &                No &        Yes &                        TOU &           No \\
      \cite{DeConinck2016a} &  DSM &         Commercial &         Simulation &                     MPC &                   &     Reduced Order &                    SS (RC) &                No &        Yes &                        TOU &           No \\
      \cite{DeConinck2016} &            Cost Savings &         Commercial &               Real &                     MPC &               Temp$^2$ &     Reduced Order &                    SS (RC) &                No &        Yes &                        TOU &          Yes \\
            \cite{Dong2014} &       Energy Efficiency &         Commercial &               Real &                     MPC &                   &     Reduced Order &                    SS (RC) &                No &        Yes &                       None &          Yes \\
          \cite{Drgona2018} &        MPC Alternatives &         Commercial &         Simulation &              Approx MPC &               Temp$^2$ &         Black Box &             NN &               Yes &        Yes &                       None &           No \\
          \cite{Drgona2019} &        MPC Alternatives &         Commercial &  Real + Simulation &              Approx MPC &               Temp$^2$ &         Black Box &             NN &                No &        Yes &                       None &          Yes \\
        \cite{Ferreira2012} &       Energy Efficiency &      Institutional &               Real &              Approx MPC &                  PMV &         Black Box &             NN &                No &        Yes &                       None &           No \\
           \cite{Finck2017} &  DSM &         Commercial &         Simulation &                     MPC &                   &     Reduced Order &                    SS (RC) &                No &        Yes &                  Day-ahead &          Yes \\
         \cite{Garnier2015} &       Energy Efficiency &         Commercial &         Simulation &                     MPC &                   &         Black Box &             NN &                No &        Yes &                       None &          Yes \\
         \cite{Gwerder2013} &            Cost Savings &         Commercial &  Real + Simulation &                     RBC &                   &                &                         &                No &        Yes &                        TOU &           No \\
             \cite{Hao2012} &  DSM &         Commercial &         Simulation &                     MPC &                   &     Reduced Order &                         SS &                No &         No &                 Regulation &           No \\
        \cite{Hilliard2017} &       Energy Efficiency &         Commercial &               Real &              Approx MPC &                   &         Black Box &                 Tree based &                No &        Yes &                       None &           No \\
              \cite{Hu2019} &            Cost Savings &        Residential &         Simulation &                     MPC &                   &     Reduced Order &                    SS (RC) &                No &        Yes &                    Dynamic &          Yes \\
           \cite{Huang2015} &            Cost Savings &         Commercial &  Real + Simulation &              Hybrid MPC &                   &     Reduced Order &                    SS (RC) &                No &        Yes &                        TOU &          Yes \\
            \cite{Jain2018} &        MPC Alternatives &         Commercial &         Simulation &                     MPC &                   &         Black Box &           Gaussian Process &                No &        Yes &                 Regulation &          Yes \\
            \cite{Jain2016} &        MPC Alternatives &        Single Zone &         Simulation &                     MPC &               Temp$^2$ &         Black Box &                 Tree based &                No &        Yes &                       None &          Yes \\
            \cite{Jain2018a} &  DSM &         Commercial &         Simulation &                     MPC &                   &         Black Box &           Gaussian Process &                No &        Yes &                 Regulation &          Yes \\
            \cite{Jain2017} &        MPC Alternatives &         Commercial &         Simulation &                     MPC &                 Temp &         Black Box &                 Tree based &                No &        Yes &                 Regulation &          Yes \\
            \cite{Jain2018b} &  DSM &         Commercial &         Simulation &                     MPC &               Temp$^2$ &         Black Box &                 Tree based &                No &        Yes &                       None &          Yes \\
             \cite{Joe2019} &            Cost Savings &         Commercial &               Real &                     MPC &                   &     Reduced Order &                    SS (RC) &                No &        Yes &                        TOU &          Yes \\
        \cite{Kahraman2018} &  DSM &         Commercial &         Simulation &                     MPC &                   &     Reduced Order &                    SS (RC) &                No &        Yes &                       None &           No \\
            \cite{Karg2018} &        MPC Alternatives &        Single Zone &         Simulation &              Approx MPC &                   &         Black Box &              Deep Learning &                No &        Yes &                       None &           No \\
         \cite{Killian2018} &            Cost Savings &         Commercial &               Real &                     MPC &                   &         Black Box &                   TS-Fuzzy &                No &        Yes &                       None &          Yes \\
           \cite{Klein2017} &  DSM &         Commercial &         Simulation &                     RBC &                   &     Reduced Order &                    SS (RC) &                No &        Yes &   Grid Support Coefficient &          Yes \\
         \cite{Knudsen2016} &  DSM &        Single Zone &         Simulation &                     MPC &                   &     Reduced Order &                         SS &                No &        Yes &    Dynamic + CO2 intensity &           No \\
          \cite{Kontes2018} &        MPC Alternatives &         Commercial &         Simulation &  RL &                   &         Black Box &           Gaussian Process &                No &        Yes &                       None &          Yes \\
             \cite{Lee2018} &           Co-simulation &         Commercial &         Simulation &                     MPC &               Temp$^2$ &     Reduced Order &                    SS (RC) &                No &        Yes &                       None &          Yes \\
             \cite{Lee2019} &       Energy Efficiency &         Commercial &               Real &                     MPC &                 Temp &     Reduced Order &                    SS (RC) &                No &        Yes &                       None &          Yes \\
         \cite{Lehmann2013} &         MPC Development &        Single Zone &         Simulation &                     MPC &                   &     Reduced Order &                    SS (RC) &                No &        Yes &                       None &          Yes \\
              \cite{Li2018} &            Cost Savings &         Commercial &         Simulation &                     MPC &               Temp$^2$ &         Black Box &                 Tree based &               Yes &        Yes &            TOU + Day-ahead &          Yes \\
             \cite{Liu2018} &  DSM &        Residential &         Simulation &                     MPC &                   &     Reduced Order &                    SS (RC) &                No &        Yes &                    Dynamic &          Yes \\
            \cite{Luzi2019} &         MPC Development &        Single Zone &         Simulation &                     MPC &               Temp$^2$ &     Reduced Order &                    SS (RC) &                No &        Yes &                       None &           No \\
        \cite{Maasoumy2014} &  DSM &         Commercial &               Real &                     MPC &                   &     Reduced Order &                    SS (RC) &                No &        Yes &                         DR &           No \\
        \cite{Maasoumy2011} &       Energy Efficiency &         Multi Zone &         Simulation &                     MPC &               Temp$^2$ &     Reduced Order &                    SS (RC) &                No &        Yes &                       None &           No \\
       \cite{Macarulla2017} &       Energy Efficiency &         Commercial &               Real &                     RBC &                   &         Black Box &             NN &                No &        Yes &                       None &          Yes \\
             \cite{Mai2015} &  DSM &   Commercial Group &         Simulation &                     MPC &                   &     Reduced Order &                    SS (RC) &                No &        Yes &                 Regulation &          Yes \\
       \cite{Manandhar2015} &         Thermal Comfort &         Commercial &                 &                     RBC &                 Temp &  Regression Based &           Linear Programme &                No &        Yes &                  Day-ahead &           No \\
       \cite{Marvuglia2014} &        MPC Alternatives &         Commercial &               Real &             Fuzzy Logic &                   &         Black Box &             NN &                No &        Yes &                       None &           No \\
            \cite{Masy2015} &  DSM &        Residential &         Simulation &                     MPC &                 Temp &     Reduced Order &                    SS (RC) &                No &        Yes &            TOU + Day-ahead &          Yes \\
      \cite{Mirakhorli2017} &            Cost Savings &        Residential &         Simulation &                     MPC &                   &     Reduced Order &                    SS (RC) &                No &        Yes &  Day-ahead + Dynamic + TOU &          Yes \\
      \cite{Mirakhorli2018} &  DSM &  Residential Group &         Simulation &                     MPC &                   &     Reduced Order &                    SS (RC) &                No &        Yes &                    Dynamic &          Yes \\
          \cite{Mocanu2019} &  DSM &  Residential Group &         Simulation &  RL &                   &        Model-free &                 Model-free &                No &         No &                        TOU &           No \\
          \cite{Nghiem2017} &        MPC Alternatives &         Commercial &         Simulation &                     MPC &                   &         Black Box &           Gaussian Process &               Yes &        Yes &                 Regulation &          Yes \\
           \cite{Ogata2019} &            Cost Savings &        Residential &         Simulation &                     MPC &               Temp$^2$ &     Reduced Order &                         SS &                No &        Yes &                        TOU &          Yes \\
      \cite{Oldewurtel2010} &  DSM &        Single Zone &         Simulation &                     MPC &                   &     Reduced Order &                    SS (RC) &                No &        Yes &                    Dynamic &          Yes \\
      \cite{Oldewurtel2012} &        MPC Alternatives &        Single Zone &         Simulation &                     MPC &                   &     Reduced Order &                    SS (RC) &                No &        Yes &                       None &          Yes \\
      \cite{Oldewurtel2013} &  DSM &         Commercial &               Real &                     MPC &                   &     Reduced Order &                    SS (RC) &                No &        Yes &                        TOU &          Yes \\
      \cite{Oldewurtel2011} &  DSM &        Single Zone &         Simulation &                     MPC &                   &     Reduced Order &                    SS (RC) &                No &        Yes &              TOU + Dynamic &          Yes \\
          \cite{Pavlak2014} &  DSM &         Commercial &         Simulation &                     MPC &                   &     Reduced Order &                    SS (RC) &                No &        Yes &                         DR &          Yes \\
            \cite{Pean2018} &  DSM &        Residential &         Simulation &                     MPC &                   &     Reduced Order &                    SS (RC) &                No &        Yes &                    Dynamic &          Yes \\
          \cite{Pippia2019} &        MPC Alternatives &         Commercial &         Simulation &                     MPC &               Temp$^2$ &     Reduced Order &                    SS (RC) &                No &        Yes &                       None &          Yes \\
   \cite{Radhakrishnan2016} &       Energy Efficiency &         Commercial &         Simulation &                     MPC &                   &     Reduced Order &                         SS &                No &        Yes &                       None &          Yes \\
         \cite{Razmara2015} &          Exergy Control &      Institutional &         Simulation &                     MPC &                   &     Reduced Order &                    SS (RC) &                No &        Yes &                       None &          Yes \\
      \cite{Reynolds2018} &            Cost Savings &         Commercial &         Simulation &                     MPC &                   &         Black Box &             NN &                No &        Yes &                        TOU &          Yes \\
         \cite{Risbeck2018} &        MPC Alternatives &         Commercial &         Simulation &                     MPC &                   &     Reduced Order &                         SS &                No &        Yes &                         DR &          Yes \\
       \cite{Robillart2019} &            Cost Savings &        Single Zone &         Simulation &                     MPC &                   &     Reduced Order &                         SS &                No &        Yes &                        TOU &          Yes \\
           \cite{Ruano2016} &            Cost Savings &      Institutional &               Real &                     MPC &                   &         Black Box &             NN &                No &         No &                        TOU &          Yes \\
         \cite{Ruelens2015} &       Energy Efficiency &        Residential &         Simulation &  RL &               Binary &        Model-free &                 Model-free &               Yes &         No &                       None &           No \\
         \cite{Salakij2016} &       Energy Efficiency &        Residential &         Simulation &                     MPC &  Temp$^2$ + Humidity$^2$ &     Reduced Order &                    SS (RC) &                No &        Yes &                       None &           No \\
         \cite{Sampaio2019} &       Energy Efficiency &        Single Zone &               Real &                     MPC &                   &     Reduced Order &                    SS (RC) &                No &        Yes &                       None &           No \\
           \cite{Shang2019} &        MPC Alternatives &        Single Zone &         Simulation &                     MPC &                   &     Reduced Order &                    SS (RC) &                No &        Yes &                       None &           No \\
             \cite{Shi2017} &       Energy Efficiency &        Single Zone &         Simulation &                     MPC &                   &     Reduced Order &                    SS (RC) &                No &        Yes &                       None &          Yes \\
          \cite{Siroky2011} &       Energy Efficiency &      Institutional &               Real &                     MPC &                   &     Reduced Order &                    SS (RC) &                No &        Yes &                       None &           No \\
          \cite{Smarra2018} &        MPC Alternatives &        Single Zone &         Simulation &                     MPC &               Temp$^2$ &         Black Box &                 Tree based &                No &        Yes &                       None &          Yes \\
          \cite{Smarra2018a} &        MPC Alternatives &        Single Zone &         Simulation &                     MPC &               Temp$^2$ &         Black Box &                 Tree based &                No &        Yes &                       None &          Yes \\
          \cite{Smarra2018a} &  DSM &         Commercial &         Simulation &                     MPC &                   &         Black Box &                 Tree based &                No &        Yes &                 Regulation &          Yes \\
          \cite{Smarra2018a} &       Energy Efficiency &        Residential &               Real &                     MPC &                 Temp &         Black Box &                 Tree based &                No &        Yes &                       None &          Yes \\
    \cite{Sturzenegger2013} &       Energy Efficiency &         Commercial &               Real &                     MPC &                   &     Reduced Order &                    SS (RC) &                No &        Yes &                       None &          Yes \\
    \cite{Sturzenegger2016} &         MPC Development &         Commercial &               Real &                     MPC &                   &     Reduced Order &                    SS (RC) &                No &        Yes &                        TOU &          Yes \\
 \cite{Tabares-Velasco2019} &            Cost Savings &        Residential &         Simulation &                     MPC &                 Temp &  Regression Based &                        ARX &                No &        Yes &                        TOU &          Yes \\
            \cite{Tang2019} &  DSM &         Commercial &         Simulation &                     MPC &                   &     Reduced Order &                    SS (RC) &                No &        Yes &                 Regulation &          Yes \\
       \cite{BinTariq2019} &         Thermal Comfort &        Single Zone &               Real &                     MPC &               Temp$^2$ &     Reduced Order &                         SS &                No &        Yes &                       None &          Yes \\
       \cite{Touretzky2014} &         MPC Development &        Single Zone &         Simulation &                     MPC &                   &     Reduced Order &                         SS &                No &        Yes &                        TOU &           No \\
      \cite{Valenzuela2019} &         MPC Development &        Single Zone &               Real &                     MPC &                   &     Reduced Order &                         SS &                No &        Yes &                       None &          Yes \\
           \cite{Vande2014} &         MPC Development &        Single Zone &         Simulation &                     MPC &               Temp$^2$ &     Reduced Order &                    SS (RC) &                No &        Yes &                       None &          Yes \\
        \cite{Vanhoudt2017} &  DSM &  Residential Group &         Simulation &                     MPC &                   &     Reduced Order &                    SS (RC) &                No &        Yes &                       None &           No \\
 \cite{Vazquez-Canteli2018} &        MPC Alternatives &        Residential &         Simulation &  RL &                   &        Model-free &                 Model-free &                No &         No &                       None &           No \\
            \cite{Viot2018} &       Energy Efficiency &        Residential &               Real &                     MPC &               Temp$^2$ &     Reduced Order &                         SS &                No &        Yes &                       None &          Yes \\
         \cite{Vrettos2016} &  DSM &   Commercial Group &         Simulation &                     MPC &                   &     Reduced Order &                    SS (RC) &                No &        Yes &                 Regulation &          Yes \\
         \cite{Vrettos2016a} &  DSM &        Single Zone &  Real + Simulation &                     MPC &               Temp$^2$ &     Reduced Order &                    SS (RC) &                No &        Yes &  TOU + Day-ahead + Dynamic &          Yes \\
         \cite{Vrettos2016b} &  DSM &   Commercial Group &         Simulation &                     MPC &                   &     Reduced Order &                    SS (RC) &                No &        Yes &                 Regulation &          Yes \\
         \cite{Vrettos2016e} &  DSM &         Commercial &               Real &                     MPC &                   &     Reduced Order &                    SS (RC) &                No &        Yes &                       None &          Yes \\
         \cite{Vrettos2013} &  DSM &        Single Zone &         Simulation &                     MPC &               Temp$^2$ &     Reduced Order &                    SS (RC) &                No &        Yes &  TOU + Day-ahead + Dynamic &          Yes \\
            \cite{Wang2019} &        MPC Alternatives &        Single Zone &         Simulation &                     MPC &               Temp$^2$ &     Reduced Order &                    SS (RC) &                No &        Yes &                       None &           No \\
            \cite{Wang2019} &        MPC Alternatives &         Commercial &         Simulation &                     MPC &               Temp$^2$ &         Black Box &                   Multiple &                No &        Yes &                       None &           No \\
             \cite{Wei2018} &        MPC Alternatives &        Single Zone &         Simulation &                     MPC &                   &     Reduced Order &                    SS (RC) &                No &        Yes &                       None &          Yes \\
             \cite{Wei2018} &        MPC Alternatives &        Single Zone &         Simulation &  RL &                   &        Model-free &                 Model-free &                No &        Yes &                       None &           No \\
             \cite{Wei2017} &        MPC Alternatives &         Multi Zone &         Simulation &  RL &                   &        Model-free &                 Model-free &                No &        Yes &                        TOU &           No \\
          \cite{Wolisz2019} &       Energy Efficiency &        Single Zone &         Simulation &                     MPC &                 Temp &         Black Box &                        ARX &                No &        Yes &                        PEF &          Yes \\
              \cite{Xu2012} &        MPC Alternatives &        Single Zone &         Simulation &                     MPC &                   &         Black Box &             NN &               Yes &         No &                       None &          Yes \\
            \cite{Zeng2015} &       Energy Efficiency &         Commercial &               Real &                     MPC &                   &         Black Box &             NN &               Yes &        Yes &                       None &           No \\
           \cite{Zhang2018} &       Energy Efficiency &      Institutional &         Simulation &                     MPC &                 Temp &     Reduced Order &                    SS (RC) &                No &        Yes &                       None &          Yes \\
           \cite{Zhang2015} &  DSM &         Commercial &         Simulation &                     MPC &                   &     Reduced Order &                    SS (RC) &                No &        Yes &                 Regulation &          Yes \\
           \cite{Zhang2013} &        MPC Alternatives &         Commercial &         Simulation &                     MPC &                   &     Reduced Order &                    SS (RC) &                No &        Yes &                       None &          Yes \\
           \cite{Zhang2019} &       Energy Efficiency &         Commercial &               Real &  RL &               Temp$^2$ &        Model-free &                 Model-free &                No &         No &                       None &           No \\
            \cite{Zhou2016} &        MPC Alternatives &         Commercial &         Simulation &                     MPC &                   &  Regression Based &  semiparametric regression &                No &        Yes &                       None &          Yes \\
            \cite{Zhou2016} &        MPC Alternatives &         Commercial &         Simulation &                     MPC &                   &     Reduced Order &                    SS (RC) &                No &        Yes &                       None &          Yes \\
            \cite{Zhou2019} &        MPC Alternatives &        Residential &         Simulation &                     MPC &                   &         Black Box &                   Multiple &                No &         No &                    Dynamic &           No \\
          \cite{Zhuang2018} &         MPC Development &         Commercial &         Simulation &                     MPC &                   &     Reduced Order &                    SS (RC) &                No &        Yes &                       None &          Yes \\
            \cite{Zong2017} &         MPC Development &        Residential &               Real &                     MPC &                   &     Reduced Order &                    SS (RC) &                No &        Yes &                  Day-ahead &           No \\
\end{longtable}

}
\end{landscape}

\subsection{Application of Studies}
\label{sec4.2}
Figure \ref{fig:1}(a) shows the categorisation of the studies based on the application of the work. The review shows that data-driven predictive control techniques have been implemented for DSM applications in a significant number of studies, justifying the need for this review. Such techniques have also been used commonly in the related applications of energy efficiency and cost savings. Significant research has also gone into alternatives of traditional model-based MPC for predictive control given some of the challenges of model and control development.   

\subsection{Building Type}
\label{sec4.3}

\begin{figure}[!htbp]
\centering
\includegraphics[scale=0.5]{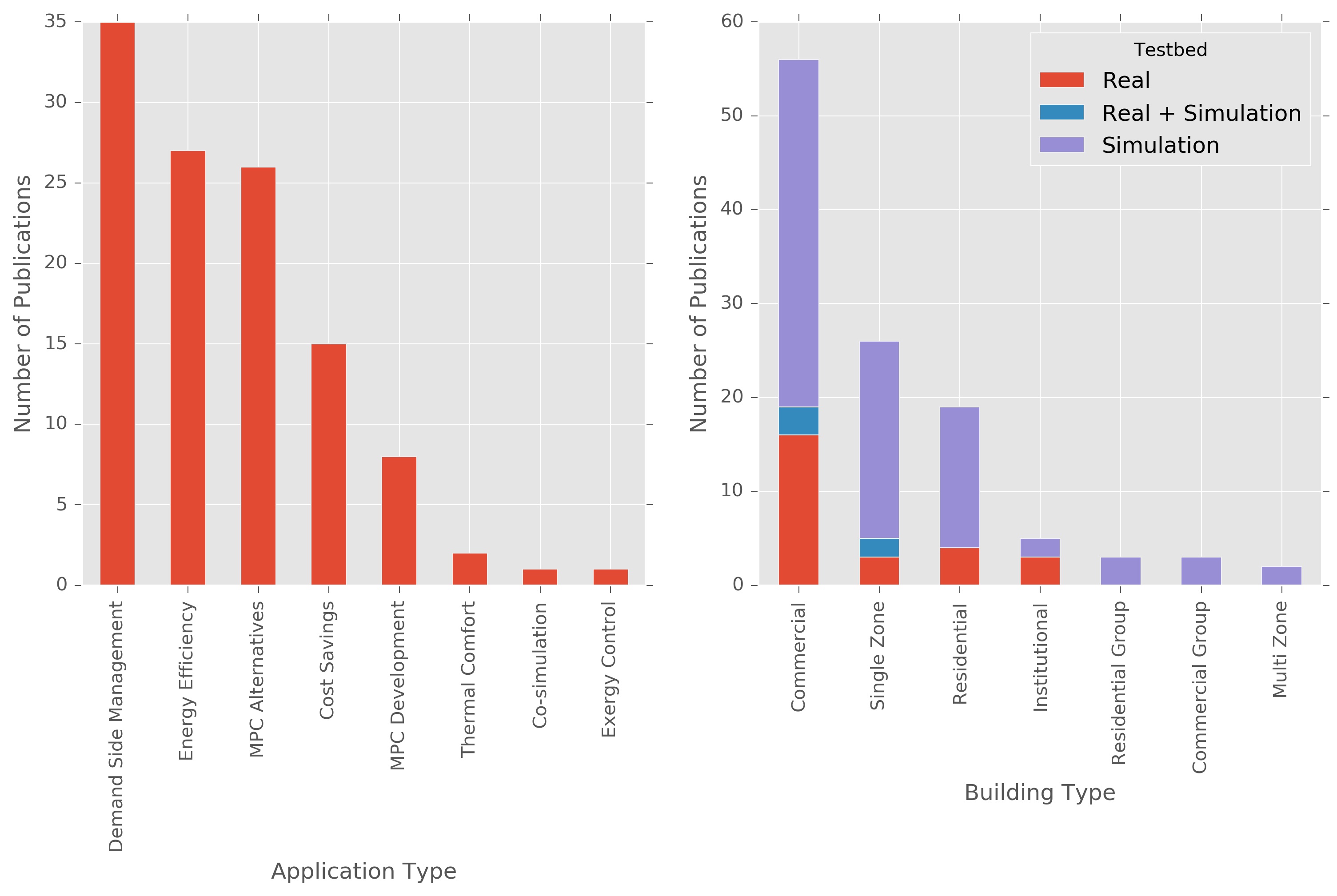}

\caption{(a) Number of Publications by Application Type, (b) Number of Publications with Building Type and Testbed Type.}
\label{fig:1}
\end{figure}

As Figure \ref{fig:1}(b) shows, commercial buildings have received significantly more attention than residential buildings in the domain of predictive control. Commercial buildings are, generally, more complex in terms of the geometry, size and energy systems present, compared to residential buildings. However, it is generally easier to obtain training data from commercial buildings given the presence of BEMS, the benefits of predictive control are more notable and the impact of occupancy is less pronounced compared to residential buildings. Single zone buildings have received a lot of attention (note that many of the studies categorised as residential are in-fact only single zones) as opposed to multi-zone buildings, which present further complexities and challenges in the modelling, e.g., capturing the thermal interactions between zones \cite{Afroz2017}. Further, many studies have approximated a multi-zone building as a single zone (e.g., through averaging zonal temperatures) \cite{Zhou2016, DeConinck2016a}. Very significantly, a small proportion of the studies were implemented on more than one building. There are some studies that have considered multiple buildings at the aggregator level but in the case of \cite{Iria2019}, they did not provide details on the modelling of the building dynamics.

\subsection{Building Thermal Dynamics Model}
\label{sec4.4}

Figure \ref{fig:2}(b) shows the publications categorised by the type of model used to simulate the thermal dynamics of the building passive thermal mass in studies where MPC is implemented (the majority in the review as shown by Figure \ref{fig:2}(a)). By far and large, the majority of studies have employed reduced order models with state-space implementation the most common type of model within this category. One explanation for this is that the theory around state-space models is well established and they are commonly used in MPC applications in many industries, and in particular the process industry.

Black-box approaches aimed at capturing the building thermal dynamics are seen to be increasing in number over the last few years. These include: tree based methods, NN, guassian processes and deep learning approaches (in decreasing order of occurrence). Generally highly nonlinear and implicitly relating the system inputs to the outputs, various approaches have been employed in literature to integrate these models as part of an optimal control framework. Some of the approaches reviewed found ways to keep the optimisation problem linear, e.g., through the separation of variables technique \cite{Behl2016, Jain2016, Jain2018, Bunning2019}, whereas others used optimisers that are able to deal with the non-linearity, e.g., Genetic Algorithms (GA) \cite{Ferreira2012, Reynolds2018} and Branch and Bound \cite{Ruano2016}, which use the data-driven models to create a discrete search space with heuristic optimisation to find the optimal solution.

One of the findings that can be made from this review is that most studies do not quantify the predictive power of the models over the length of prediction horizon (as \citet{Privara2013} did). This is not such an issue for receding horizon approaches (such as MPC), however, the impact on performance of predictive control is not well known. For certain black-box models (e.g., NN and tree-based models), model pruning is also an essential step to ensure that the model does not overfit (i.e., generalises poorly on unseen data).

\subsection{Control Approaches}
\label{sec4.5}

\begin{figure}[!htbp]
\centering
\includegraphics[scale=0.5]{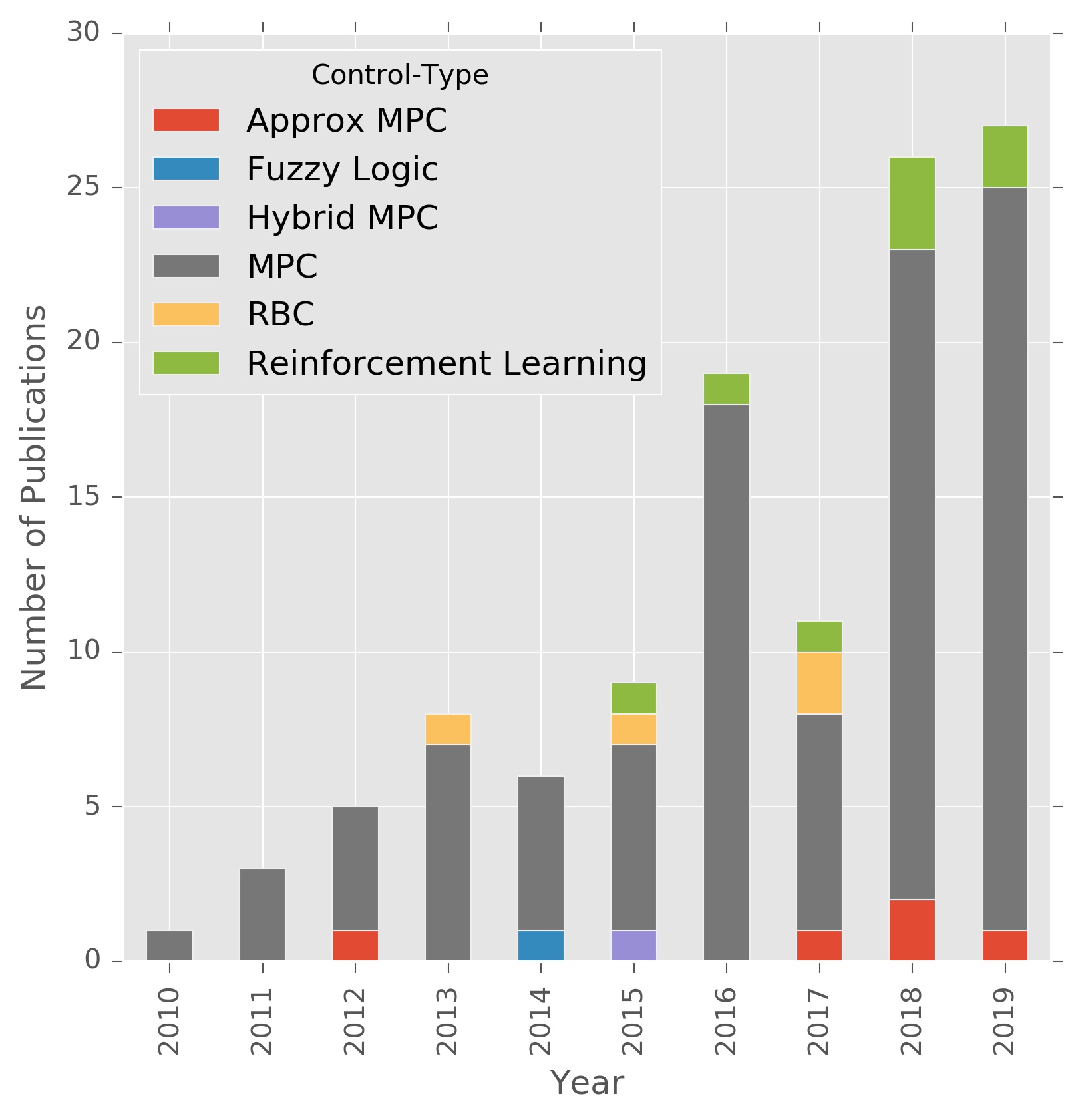}
\includegraphics[scale=0.5]{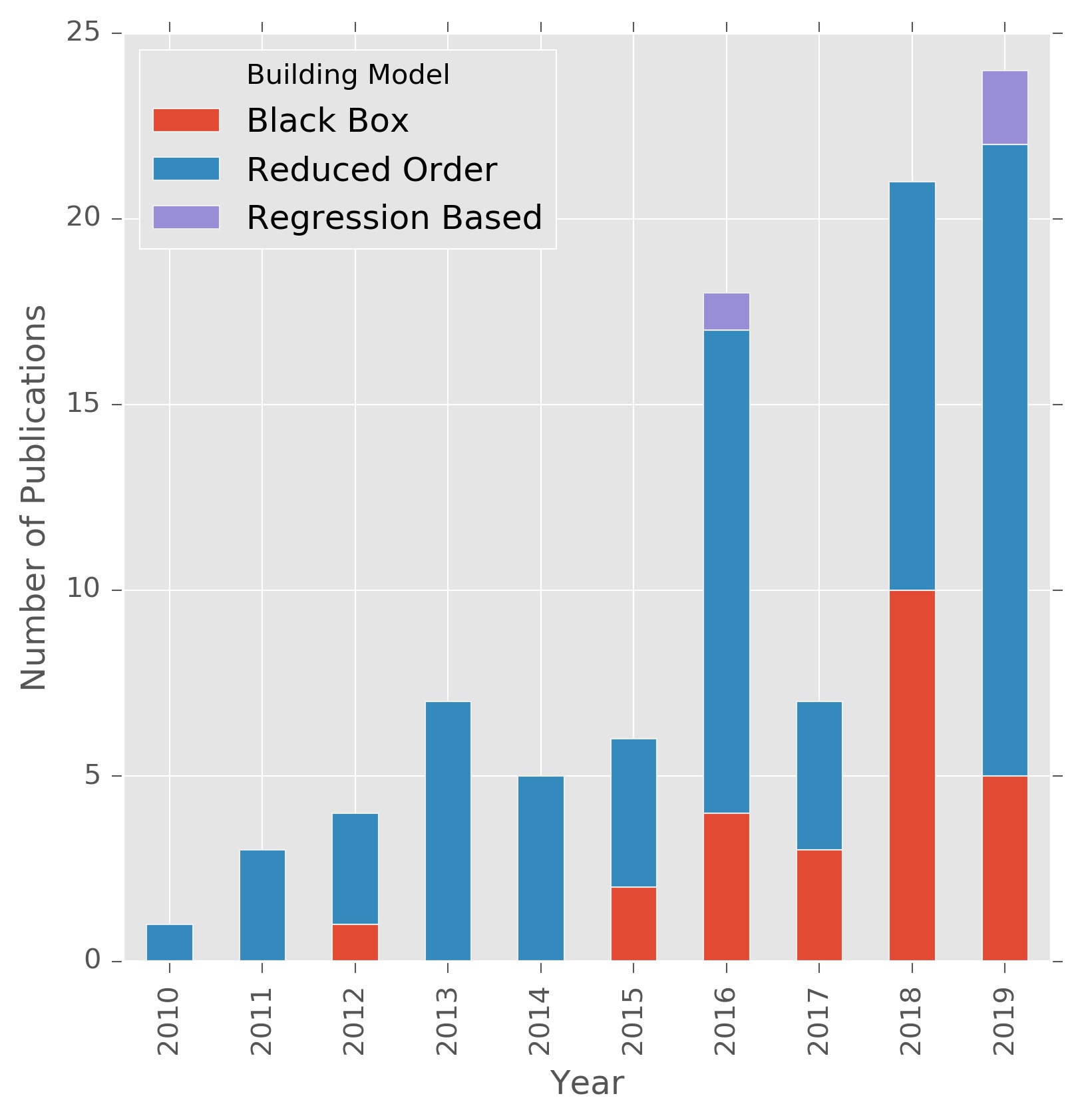}

\caption{(a) Number of Publications by Year and Control Type, (b) Number of MPC Publications by Year and Building Model Type.}
\label{fig:2}
\end{figure}

Figure \ref{fig:2}(a) shows the number of publications per year categorised by the type of control used. The first major trend that is visible is the increasing interest and publication of research in the area of the data-driven predictive control of building energy systems. Whilst MPC approaches dominate the studies, the use of black-box model based MPC shows an upwards trend in more recent years (Figure \ref{fig:2}(b)). RL as a model-free control technique has seen increased interest in the last few years and this is also seen in the more comprehensive review of the technique by \citet{Vazquez-Canteli2019}. The other control techniques used in the reviewed papers had small sample sizes.

One domain that has attracted increased attention is data-driven approaches that learn and approximate based on MPC input-output training data \cite{Drgona2018, Drgona2019, Ferreira2012, Hilliard2017, Karg2018}. \citet{Drgona2018, Drgona2019} used Deep Time Delay Neural Network (TDNN) to train on MPC data based on a physics-based model and was able to reduce the computational and memory requirements significantly using this model instead. The authors argue that such an approach has the benefits of a low computational footprint, minimal software dependencies and easy deployment on low level hardware. Whilst this approach has benefits from an operational standpoint, the need for development of the original MPC model and controller that provide the training data provides challenges for implementation.

\subsection{Optimisation Objectives}
\label{sec4.6}

The optimisation objective function ultimately determines the global objectives desired from the controlled process and are usually determined by the stakeholders, in this case, typically the building owner or energy manager. The preferences of the occupants of conditioned spaces may also influence the objectives of the optimisation. These objectives are usually related to thermal comfort (e.g., maintaining internal zone temperature within bounds and minimising occupant discomfort hours), economic goals (e.g., minimising overall energy demand or operational costs), or environmental goals (e.g., maximising RES self-consumption or reducing CO$_2$ intensity of energy consumed). Figure \ref{fig:3}(a) plots the number of publications categorised by the objective function considered. Minimisation of energy demand is the most commonly considered objective function in the studies reviewed, followed by minimisation of cost (economic) and minimisation of discomfort. Note that time-varying price signals are required for a cost objective function to be relevant. Given this, however, a day-ahead price schedule or forecast for RTP is all that is required to change an energy minimisation problem into a cost minimisation problem. An economic objective function would also allow the capture of revenue from DR programmes to be quantified in the case of incentive based DR. The effectiveness of the pricing signal is crucial in enabling and promoting the energy flexibility of a building to be utilised. Note that the use of an economic cost objective function may not necessarily lead to the best environmental outcome as, firstly, such a control objective may incentivise the use of greater overall energy consumption, and second, the pricing signal may not correlate well with the CO$_2$ emissions from generation.

Comfort considerations are taken into account in the reviewed literature predominantly through quantifying the deviation of internal temperatures (either state or output variable) from either a setpoint or allowable band of temperatures. These are very often time-varying and can either be found in the objective function or in the constraints (either as hard or soft versions with a weighted slack variable in the objective function). Very few studies considered other thermal comfort parameters such as Predicted Mean Vote (PMV), which directly describes user thermal comfort and hence are not as conservative as temperature control \cite{Cigler2012}. However, this does pose additional challenges such as resulting in a nonlinear optimisation problem increasing computational complexity and requiring further knowledge of occupants such as activity and clothing levels.  

Investigating the form of the objective function, it can be seen that for the comfort objectives, quadratic cost functions are the most common. This is representative of tracking problems (e.g., tracking the temperature setpoint) where the quadratic form helps with stability and reduced computational effort of the optimisation \cite{Serale2018}. Otherwise, linear functions are used, and where the problem is multi-objective (e.g., minimising discomfort and minimising operational cost), appropriate weighting terms must be given to the different objectives based on the stakeholder requirements and priorities.

Other objectives found in the reviewed literature include stability of the controller \cite{Aoun2019}, minimising PV curtailment \cite{Coetzee2019}, minimising peak loads \cite{Jain2018, Kahraman2018}, minimising CO$_2$ emissions \cite{Knudsen2016}, minimising exergy destruction \cite{Razmara2015}, minimising the primary energy factor \cite{Wolisz2019} and provision of reserve \cite{Vrettos2016e, Vrettos2016} although these are very much in the minority. However, this also shows the flexibility of the MPC approach to differing stakeholder requirements hence its increased interest in the building energy management area. The stability requirement of the controller for building energy management is generally relaxed due to the slower dynamics found in buildings compared to other systems and hence control objectives are dictated by performance objectives primarily \cite{Oldewurtel2012}.

\subsection{Energy Flexibility Resources Considered}
\label{sec4.7}

\begin{figure}[!htbp]
\centering

\includegraphics[scale=0.5]{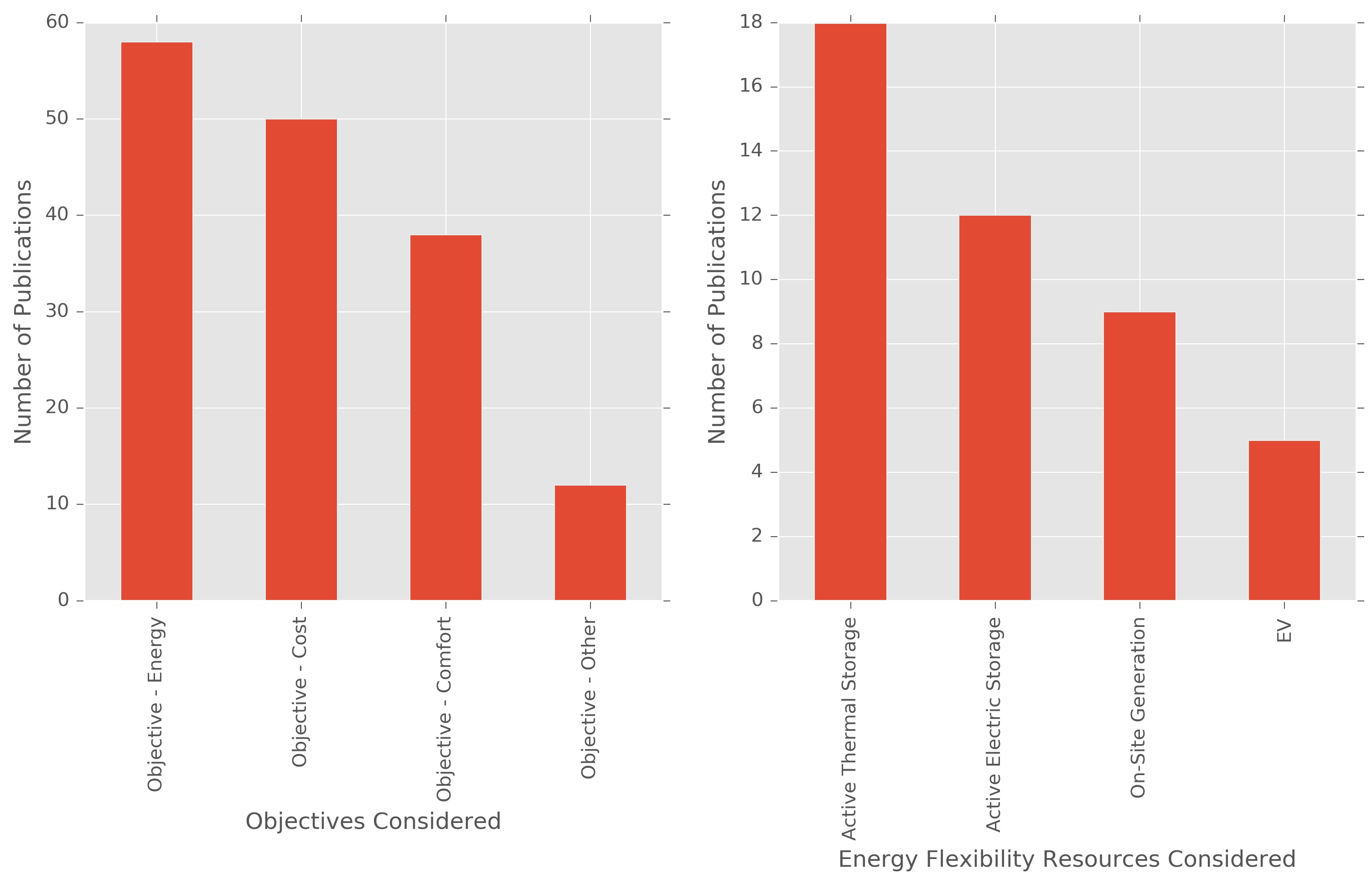}

\caption{(a) Number of Publications by Optimisation Objectives Considered, (b) Number of Publications by Energy Flexibility Resources Considered (excluding building passive thermal mass).}
\label{fig:3}
\end{figure}

A key objective of this review is to investigate the energy flexibility resources considered in the studies utilising predictive control. All studies reviewed with an application of DSM here utilised building passive thermal mass. Building passive thermal mass receives the most attention due to the significant potential it has for energy flexibility applications, but also for the challenges mentioned previously regarding suitably modelling the thermal-dynamics of buildings for control applications. When considering other energy flexibility sources, as Figure \ref{fig:3}(b) shows, considerably fewer studies considered active thermal energy storage systems and active electrical energy storage systems, such as batteries. Even fewer studies considered EVs and on-site generation as flexibility resources within the predictive control framework. Only one study considered all of the flexibility resources mentioned above \cite{Mirakhorli2018}. The study by Mirakhorli and Dong considered 15,000 residential buildings integrating the grid and nodal pricing to coordinate the operation of the buildings.

\subsection{Feature Selection}
\label{sec4.8}
Considering the studies in the review, only six employed a formalised feature selection technique (i.e., a process in the model development pipeline for reducing the dimension of the input space to the data-driven model) as part of their methodology. \citet{Drgona2018} employed a three step process in selecting the features for use in the Deep TDNN model to approximate MPC laws. The steps involved are a manual engineering judgement based on elimination of linearly dependent features (i.e., those input features that are highly correlated and hence they do not add significant information to the machine learning model), a principal component analysis based dimensionality reduction and, finally, selection of the disturbance features based on the model disturbance dynamics. \citet{Li2018} used an if-then rules based optimal feature selection framework, although this was limited to the selection of lag terms based upon the day of the week. \citet{Nghiem2017} were required to use the Automatic Relevance Determination tool of Gaussian Processes (GPs) for selecting the relevant inputs for the GPs modeling the response of the buildings to a control signal. \citet{Ruelens2015} employed an ANN based auto-encoder for dimensionality reduction in the application of a model-free RL strategy for reducing the energy consumption of a heat pump. \citet{Xu2012} and \citet{Zeng2015} used the boosting tree algorithm for dimensionality reduction for neural network models. This review highlights that most studies do not consider the issue of feature assessment to be important, as \citet{Schmidt2018} also concluded in their review study, and that when employed, there is a lack of a standard methodology or process in these studies.

\subsection{Forecasts Used}
\label{sec4.9}

The studies were categorised based on the predictive inputs used for the predictive controller. Almost all the studies considered included future predictions of external weather variables as inputs (93\% of studies considered). Investigating the studies that did not consider future weather inputs, \citet{Borsche2014} developed a predictive control strategy for active thermal storage only, which if sufficiently well insulated, is somewhat damped from the effects of ambient conditions. \citet{Hao2012} assumed constant ambient conditions as part of the linearisation process for the building thermal dynamics.  \citet{Vazquez-Canteli2018} and \citet{Zhang2019} utilised RL control approaches which learn from the state-reward relationship over time and hence can be implemented without the requirement for predictive weather forecasts. However, whether there are potential performance improvements when weather forecasts are used as part of the state in the RL problem is not known and should be investigated. \citet{Zhou2019} trained data-driven models based on past data from an MPC controller with purely past inputs (ambient temperature and electricity price). In most cases, offline predictions are used (which rely on data which has already been measured) with exact forecasts of disturbances used in testing the controller. This is frequently done to simplify the implementation of the controller and performance bounds are obtained in such cases (i.e., in reality, there will always be errors in the forecasting of weather and occupancy degrading the performance of the controller with respect to its objectives). Studies have looked at the effect of uncertainties in weather predictions \cite{Bianchini2019} and pricing \cite{Borsche2013}. SMPC \cite{Shang2019, Oldewurtel2012, Zhang2015} and RMPC \cite{Zhang2013} are variations to traditional MPC that are better able to deal with uncertainties in the forecast.

\begin{figure}[!htbp]
\centering
\includegraphics[scale=0.5]{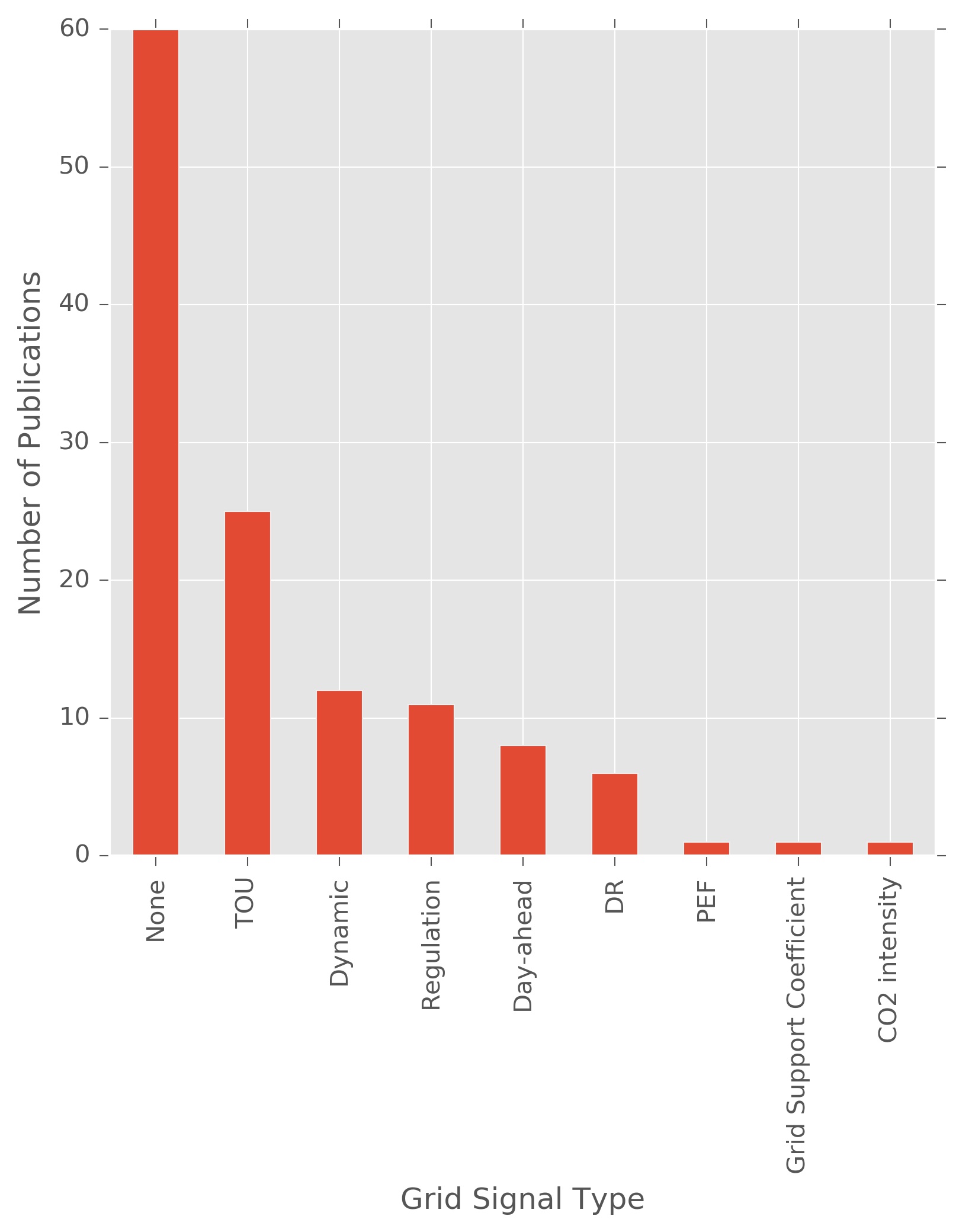}
\caption{Number of Publications by Type of Grid Signal Considered.}
\label{fig:4}
\end{figure}

The ability to receive and react to forecast energy grid signals is crucial for any DSM measure. The grid signals considered by the studies is illustrated in Figure \ref{fig:4}. Approximately 48\% of studies considered grid signals as part of the control framework. The most common pricing signal used was TOU (static) prices, which are currently very common in electricity markets and, as rates and times are fixed in advance, do not require a forecast model. Dynamic pricing signals, such as RTP and spot market prices, apply variations in pricing at a more granular level (both temporal and locational). These are the next most common and are usually easily accounted for in MPC schemes, although these require forecasting for the prediction horizon (unless exact knowledge is assumed like the majority of studies). Various techniques, such as Support Vector Regression \cite{Oldewurtel2010} have been used when forecasts of grid signals are required. For a full review of the state-of-the-art in short-term electricity price forecasting, the reader is referred to \cite{Weron2014}. 
Regulation signals are a power consumption profile for the building to follow and in this case, the objective function will be the tracking error of the building power consumption. Other grid signals considered in the reviewed papers are the grid support coefficient \cite{Klein2017} (measuring the impact of additional load based on the electricity demand on the grid at a time instance) and CO$_2$ intensity \cite{Knudsen2016}. Generally, these signals and forecasts are provided directly from the grid-side and can be considered to be a third-party input to the predictive control framework of an individual building.

Forecasting and use of occupancy as a disturbance was also considered in the majority of studies (66\%). Given that building energy systems have the primary role of satisfying occupant needs, the ability to detect whether a building or zone is occupied is fundamental to building energy management and control. Many studies assumed a constant schedule (primarily for commercial buildings where the occupancy patterns are more predictable due to given working hours). For more details on the state-of-the-art in short term predictions of occupancy in commercial buildings, the reader is referred to \citet{LiZ2018}. For residential buildings, the occupancy trends are more stochastic and random and the assumption of a constant schedule is less justifiable \cite{Buttitta2020}. \citet{Kleiminger2014} provides a comparative review of forecasting approaches for household occupancy using past records of the occupancy state. See \citet{Serale2018} for a review of prediction models of disturbances in MPC studies for energy efficiency applications.

\subsection{Validation of Approaches}
\label{sec4.10}

As Figure \ref{fig:1}(b) illustrates, most controllers are tested only on virtual (surrogate simulation) testbeds such as through the use of a co-simulation environment. In these cases, the surrogate simulation is generally a physics-based white-box model although the control-oriented building model can also be used to test the performance of the controller \cite{Serale2018}. Few studies have implemented predictive controllers on real buildings and there are numerous reasons for this. One of the reasons that this may be the case is that building owners and managers are not willing to take a risk on compromising occupant comfort through modifications or replacement to the building control system for testing purposes. This is especially true when significant hardware changes are required as part of implementing and testing the controller and the reliability of such controllers has not been proven. Further, in experimental applications, the predictive controller needs to be integrated with the BAS system of the building which can be a time-consuming and costly process. However, testing the controller on a real building presents the challenge of quantifying the savings achieved through predictive control as it is not usually possible to also test the reference control for the exact same conditions.

\subsection{Quantitative Performance of Techniques}
\label{sec4.11}
A summary of the data-driven predictive controller performance taken from 10 relevant papers is presented in a quantitative manner in Table 3. Data-driven predictive control has generally been shown to outperform baseline RBC systems, for the various controller objectives encountered. From the sampled studies, where the controller performance is quantified, an average saving of 23\% is seen for either cost or energy consumption compared to the RBC baselines. This is often due to the capabilities of data-driven predictive control to take into account the forecasts of exogenous inputs such as weather and price signals. However, it is not a trivial exercise to compare the performance of such controllers from different studies in any meaningful manner, given the different building types, exogenous disturbances (such as weather and grid signals) and time periods considered. Therefore, the efficacy of the different techniques cannot readily be assessed and compared.

\begin{table}[h!]
\centering
\caption{Comparison of Quantitative Performance of a Sample of Studies Implementing Data-Driven Predictive Control.}
\scriptsize
{\setlength\tabcolsep{3.5pt} 
\label{table:3}
\begin{tabular}{p{0.5cm}p{2cm}p{0.7cm}p{3cm}p{2cm}p{1.5cm}p{3.5cm}p{1cm}}
\toprule
                        Ref &              Author &  Year &               Model Type &            Control Type &            Testbed &           Controller Performance &              Baseline \\
\midrule
           \cite{Afram2017} &            Afram, A &  2017 &           Black-box - NN &                     MPC &         Simulation &      6-73\% saving cost &  RBC \\
         \cite{Bunning2019} &          Bunning, F &  2019 &   Black Box - Tree Based &                     MPC &               Real &        25\% saving cooling energy &      RBC \\
        \cite{Costanzo2016} &         Costanzo, G &  2016 &           Black-box - NN &  RL &  Real + Simulation &  90\% within mathematical optimum &                   MPC \\
      \cite{DeConinck2016} &       De Coninck, R &  2016 &  Reduced Order - SS (RC) &                     MPC &               Real &      30\% saving cost &                   RBC \\
         \cite{Gwerder2013} &          Gwerder, M &  2013 &               Model-free &        RBC (Predictive) &  Real + Simulation &      15\% saving cost &                   RBC \\
        \cite{Hilliard2017} &         Hilliard, T &  2017 &   Black Box - Tree Based &              Approx MPC &               Real &                29\% saving energy &                   RBC \\
           \cite{Huang2015} &            Huang, H &  2015 &  Reduced Order - SS (RC) &              Hybrid MPC &  Real + Simulation &      13\% saving cost &                   RBC \\
       \cite{Marvuglia2014} &        Marvuglia, A &  2014 &           Black-box - NN &             Fuzzy Logic &               Real &                        Not given &                   N/A \\
        \cite{Robillart2019} &        Robillart, M &  2019 &           Reduced Order - SS &             MPC &               Simulation &    6-13\% saving cost &                   RBC \\
 \cite{Tabares-Velasco2019} &  Tabares-Velasco, P &  2019 &   Regression Based - ARX &                     MPC &         Simulation &      30\% saving cost &                   RBC \\
\bottomrule
\end{tabular}

}
\end{table}
\normalsize

\section{Discussion}
\label{sec5}

With data-driven models being increasingly used for building control applications in the domain of DSM, in addition to more traditional domains such as energy efficiency and economic management, there is a need to understand the data features that are most relevant to this application. Further, given the ``Internet of Things" revolution and consequent rapid explosion of use of sensors in buildings and resulting availability of building data, feature assessment and selection is an increasingly important part of the data-driven model development ``pipeline". Only 5\% of the studies considered feature selection in model development which questions how suitable and robust the methods used in the reviewed studies would be for large complex buildings with many sensors and data features available. For example, \citet{Drgona2018} were able to use a simple feature selection approach to reduce the complexity of the controller and reduce the implementation cost of the predictive controller. Further benefits of feature selection are increasing the predictive accuracy of the data-driven models and this allows greater amounts of energy flexibility to be harnessed with lower probability of comfort and other constraints being breached. Feature selection also allows a better understanding of the underlying phenomena and processes involved in the models.

As Figure \ref{fig:2}(b) showed, there have been an increasing number of studies that have integrated a control-oriented black-box model for the process dynamics with a MPC controller (DDPC or DPC). Given that black-box models are often highly non-linear, this poses challenges for integrating with the optimisation framework present in MPC. Some of the approaches reviewed found ways to keep the optimisation problem linear whereas others used optimisers that are able to deal with the non-linearity, albeit at a computational cost. Depending on the length of the control timestep, this may or may not be an issue. Each approach has its own peculiarities and benefits, e.g., the ``separation of variables" technique ensures a convex optimisation problem with a guaranteed solution; the Branch and Bound technique provides the advantages of a guaranteed optimal solution and poses no constraints on the formulation of the objective function although it is a heuristic optimisation method. As explained further in the next section, it is challenging to make meaningful comparisons of these different techniques given the reviewed studies and no benchmarking.

A high level of simplification can be observed in the studies reviewed, highlighting the challenges faced in reality when modelling the building stock. Many studies approximated multi-zone buildings as a single-zone for the building thermal dynamic model. Given that zones may have very different thermal behaviours (given differing occupancy patterns and energy gains), especially in commercial building cases, the implications of such simplifications are not always fully understood or quantified. There have also been very few studies that have considered clusters or groups of buildings in the implementation of predictive control. This is further compounded by the fact that most studies perform validation of the model and control on surrogate simulations, as outlined in Section \ref{sec4.3} (see Figure \ref{fig:1}b). Fewer studies test the controller in a real building. This raises the question whether the findings of one case study for a given building are replicable given another building, as well as questioning the scalability and generalisability of the studies and methods in the literature. These findings are also identical in the problem domain of building energy consumption prediction, with the majority of studies testing a machine learning modeling framework on data from a single building \cite{Miller2019}. Often, the studies offer an improvement that would be dwarfed by the level of variance that would be seen when employing the algorithm on many real-world datasets. As \citet{Miller2019} points out, there is a lack of benchmarking over many buildings (both the datasets and processes) for prediction based on meter data and the same applies to the domain of predictive control. This lack of benchmarking is also reflected when attempting to compare the efficiency and value of the existing strategies reviewed as part of this study (see Table 3). There is little consistency nor standard benchmarks available for the training data, assumed boundary conditions, grid signals, market parameters or buildings investigated. Whilst such environments do exist in certain specific domains, e.g., the implementation of RL agents for the district level building DSM problem (CityLearn \cite{Vazquez-Canteli2019a}), more general purpose environments that also allow and cater for other data-driven approaches are lacking.

Less than 1\% of studies considered five energy flexibility resources in the predictive control framework (building passive thermal mass, active thermal storage, active electric storage, EV and on-site generation). Over 75\% of studies considered only one resource (predominantly building passive thermal mass). Although likely representative of the current building stock in general, when only buildings with high DSM potential (either large consumption or availability of energy flexibility resources) are considered, a predictive control framework able to account for multiple-energy vectors is required. This is a pressing need given the rise in electric storage \cite{Tsiropoulos2018}, EVs \cite{Bunsen2019} and on-site generation \cite{InternationalEnergyAgencyIEA2017a, Daneshazarian2018} in both commercial and residential buildings anticipated.   

\section{Conclusions}
\label{sec6}

In the vast majority of buildings, the control loop between the grid and the building has not been closed, which has left these buildings unable to become significant participants in DR programmes and to contribute to the balancing needs of the future smart grid. Data-driven predictive control has the potential to be a framework to allow this closing of the loop, where deriving physics based control suitable models of the building is difficult or incapable of being scaled. This article reviewed 115 current implementations of data-driven predictive control in building energy management and provides the following key insights:

\begin{itemize}
    \item Data-driven predictive control has been shown in the literature to be a very promising technique for enabling grid-interactive buildings that can harness the passive thermal mass embedded within the building. The literature shows meaningful integration with the grid and carbon footprint reduction at the single building level. Model Predictive Control and variants (e.g., Robust Model Predictive Control, Stochastic Model Predictive Control) dominates as one of the control strategies with the most research interest with Reinforcement Learning garnering increased interest, particularly in recent years.
    \item Given the increased installation of sensors in buildings and availability of data features, there is little justification for the features selected and used in studies to train the building models. Moreover, no standard methodology is utilised in the model and controller development pipeline for the application of feature selection and engineering.
    \item For optimal control strategies, the building model needs to be compatible with the optimisation formulation with linear or quadratic models guaranteeing convexity. Many varied approaches for integrating black-box models with traditional Model Predictive Control with differing benefits and costs have been utilised in the literature. There have been few comparisons made in this relatively novel domain and there is a need for both benchmarking datasets, as well as processes and toolkits to facilitate benchmarking. Without these benchmarking tools, it is very difficult to compare the efficacy and scalability of the different data-driven techniques implemented to date, due to the varying buildings and dynamic boundary conditions used in individual studies.
    \item Most predictive control applications that utilise building passive thermal mass are only implemented on single buildings. The extension of such methodologies to the wider building stock remains an open question. Therefore, for the widespread adoption of predictive control for Demand Side Management, the scalability of the approaches needs to be proven. 
    \item Most predictive control frameworks are only validated and tested on surrogate simulations. The ease of implementation and integration with Building Automation Systems as well as the ability of the controller and model to react to real stochastic disturbances is not commonly considered in the studies. 
    \item Few studies that utilise building passive thermal mass have also considered other sources of building energy flexibility (e.g., thermal energy storage, batteries, electric vehicles) and onsite generation, which are increasingly being integrated with building energy systems. 
\end{itemize}

\section{Recommendations and Future Work}
\label{sec7}
Data-driven predictive control has been shown to be a promising candidate for building integration with the electrical grid and unlocking building energy flexibility. However, the results of this review indicate some research areas that require further attention to accelerate the market accessibility and penetration of these techniques. These research gaps include: feature selection, need for benchmarking, influence of data quality and scalability in the model and control development process. To address these areas of concern, the following research questions are posed as future challenges:
\begin{itemize}
    \item What is a suitable and robust methodology for selecting the data features most relevant for building energy flexibility applications? What are these data features that are essential for a given type of building?
    \item What data-driven technique is most suitable for predictive optimal control for building energy management and harnessing energy flexibility in multi-energy vector buildings? What is a suitable benchmarking dataset and process for comparing these different techniques?
    \item What is the influence of data quality (synthetic/real, training periods, etc.) on the performance of the predictive control and model and its consequent impact on harnessed energy flexibility? 
    \item How can the data-driven model and control development process for one building be modified to allow a more transferrable and scalable approach to other or multiple buildings?
\end{itemize}


\section*{Acknowledgements}
\label{sec8}

The authors gratefully acknowledge that their contribution emanated from research supported by Science Foundation Ireland under the SFI Strategic Partnership Programme Grant Number SFI/15/SPP/E3125.

\bibliography{MyCollection}

\end{document}